# Self-similar collapse in a circular magnetic field and electron beam jets by hybrid transverse plasmon

X.L. Liu[1,a] and X.Q. Li[2]

[1]Department of Physics, Nanchang University, Jiangxi, Nanchang330031, China

[2]Department of Physics, Nanjing Normal University, Nanjing210097, China

**ABSTRACT**

Based on the set of nonlinear coupling equations describing the interaction of the high-frequency field, the self-generated magnetic field and the ion-acoustic field, the dispersion relation for the circular magnetic field is obtained. The numerical results indicate that the strength of the magnetic field have influence on the growth rate of modulation instability. The self-generated magnetic field has the tendency to self-similar collapse which makes the electron escapes along the axial region and form collimated jets. The velocity of jets is calculated and the results are consistent with experimental observations. The research may be applied to understand the dynamic process of electron beam jets in laboratory and space plasma.

**Key Words**: modulation instability; the maximum growth rate; circular self-generated magnetic field; electron beam jets

[a] e-mail: xiaolanliu@ncu.edu.cn

## 1 Introduction

In recent decades, many researchers reported that megagauss magnetic fields and collimated electron jets with high-energy density in laser-plasma have been observed experimentally [1-4]. The collimated jets sources with high-energy density play an important role for numerous applications [5-7], such as laser energy deposition, laser fusion, X-ray generation, etc. The generation mechanisms of the collimated electron jets have been discussed [8,9]. And the culminating effect of self-generated magnetic fields has recently been addressed in Refs.[10,11] in a different context.

Our research showed that, originally this behavior of the self-generated magnetic fields can be described by a set of nonlinear equations (4) -(6) [12,13] [see below]. In later sections of the paper, we will show that circular magnetic field, observed in many laser plasma experiments[14,15], is a exact solution of the nonlinear equations for a equilibrium states; and, with respect to small disturbances with limited amplitude, the equilibrium states are not stable in Lyapunov sense, leading to enhancements of the electric field and the circular field in local region; the ponderomotive force from the enhanced electric field repulses a material and forms coronal area with a low density; at the same time, the contraction of the magnetic field make the material escape along the axial region and form collimated plasma beam jets.

The paper is organized as follows. In Sec. II, the dynamic set of equations for the self-generated magnetic field in laser plasma are described first. In Sec. III, the dispersion relation of circular magnetic field is derived and the numerical study of the growth rates has been obtained. Then, the high energy electron beam jets are calculated in in Sec. IV. The discussion and conclusion are presented in Sec. V.

## 2 Field equations for self-generated magnetic flux

Starting from Vlasov equation:

$$\frac{\partial f_{e,i}}{\partial t} + \mathbf{v} \cdot \frac{\partial f_{e,i}}{\partial \mathbf{r}} + \mathbf{F} \cdot \frac{\partial f_{e,i}}{\partial \mathbf{p}} = 0, (\alpha = e, i), \tag{1}$$

where $\mathbf{F}$ is the electromagnetic force, in which $\mathbf{E}$ and $\mathbf{B}$ satisfy the Maxwell equations. $f_\alpha$ is the particle distribution function:

$$\int f_\alpha \frac{d\mathbf{p}}{(2\pi)^3} = n_\alpha,$$

where $n_\alpha$ is the particle density. The distribution function and the electromagnetic field could be divided into an unperturbed and perturbed parts, $f_\alpha = f_\alpha^R + f_\alpha^T$, and the unperturbed parts of the electromagnetic field vanish. Assuming that the perturbation field $E^T$ is weak, the condition is met:

$$\overline{W}^p \equiv \frac{|\mathbf{E}^T|^2}{8\pi n_0 k_B T_0} < 1. \tag{2}$$

Then the perturbed distribution function can be expanded in powers of the perturbed field:

$$f_\alpha^T = \sum_i f_\alpha^{T(i)},$$

where the superscript $i$ in brace denotes the $i$-th order of $E^T$. Expanding $A = (\mathbf{F}^T, f_\alpha^T)$ in a Fourier series,

$$A(\mathbf{r}, \mathbf{p}, t) = \int A_k e^{-i\omega t + i\mathbf{k} \cdot \mathbf{r}} dk, k = (\mathbf{k}, \omega),$$

and taking Maxwell equations into consideration, we get the following field equations:

$$(k^2 - \frac{\omega^2}{c^2}\varepsilon_k^t)E_k^{Tt} = \frac{4\pi i}{c^2}\omega[\mathbf{e}_k^{t*} \cdot (\mathbf{j}_k^{(2)} + \mathbf{j}_k^{(3)})], \tag{3a}$$

$$\varepsilon_k^l E_k^{TS(l)} = -\frac{4\pi i}{\omega}[\mathbf{e}_k^{l*} \cdot (\mathbf{j}_k^{(2)} + \mathbf{j}_k^{(3)})], \tag{3b}$$

here $\varepsilon_k^\delta$ is the dielectric constant for $\delta$-mode ($\delta = l$, longitudinal mode, $\delta = t$, transverse mode), $\mathbf{e}_\mathbf{k}^t$ the polarization vector, and the nonlinear currents are:

$$\mathbf{j}_k^{(2)} \propto \sum_\alpha S_{k,k_1,k_2}^\alpha E_{k_1}^T E_{k_2}^T,$$

$$\mathbf{j}_k^{(3)} = \sum_\alpha G_{k,k_1,k_2,k_3}^\alpha E_{k_1}^T E_{k_2}^T E_{k_3}^T,$$

the symbol $\sum_{\alpha}$ in the expression above implies the contributions of electrons and ions. As the ion mass is by far the larger, we can neglect their contribution:

$$S^{\alpha(t)}_{k,k_1,k_2} \approx S^{e(t)}_{k,k_1,k_2}, G^{\alpha(t)}_{k,k_1,k_2,k_3} \approx G^{e(t)}_{k,k_1,k_2,k_3}.$$

Calculating the matrix elements of interactions and putting $E_k^{Tt} = E_k^{TS}$ for low frequency transverse field in Eq.(3a), we get the field equation for $\mathbf{B}_k^S (= \mathbf{E}_k^{TS} \times \mathbf{k}c/\omega)$ in Fourier space. And for the high frequency transverse field the case is similar. Moreover, we get the equation for perturbed densities $\delta n_e$ of electrons, which are caused by the "collision" of the transverse plasmons with positive and negative high frequencies, involving the effects of the first-order $f_\varepsilon^{T(1)} \propto E^{T(l)}$ and the second-order $f_\varepsilon^{T(2)}$ in the perturbed distribution function $f_\varepsilon^T$. In the coordinate representation, by the introduction of envelope for the high-frequency transverse (electric) field,

$$\mathbf{E}(\mathbf{r},t)e^{-i\omega_{pe}t} = \int \mathbf{E}_k^{T(+)} e^{-i\omega t + i\mathbf{k}\cdot\mathbf{r}} dk, \qquad (3c)$$

eventually, we can obtain a set of equations by transverse plasmon, as follows[16]:

$$(\frac{\partial^2}{\partial \tau^2} - \nabla^2)n(\tau,\xi) = \nabla^2 |\mathbf{E}(\tau,\xi)|^2, \qquad (4)$$

$$i\frac{\partial}{\partial \tau}\mathbf{E} - \alpha \nabla \times \nabla \times \mathbf{E}(\tau,\xi) = n\mathbf{E}(\tau,\xi) + i\mathbf{E}(\tau,\xi) \times \mathbf{B}(\tau,\xi), \qquad (5)$$

$$(-\frac{\partial^2}{\partial \tau^2} + \nabla \times \nabla \times)\mathbf{B}(\tau,\xi) = i\frac{2}{3}\nabla \times \nabla \times \{\frac{\partial}{\partial \tau}[\mathbf{E}(\tau,\xi) \times \mathbf{E}^*(\tau,\xi)]\}, \qquad (6)$$

with

$$\nabla \cdot \mathbf{E}(\tau,\xi) = 0, \qquad (7)$$

where $\mathbf{E}(\tau,\xi)$ is the dimensionless envelope for the high-frequency field, $\mathbf{B}(\tau,\xi)$ is the dimensionless low-frequency self-generated magnetic field, $n(\tau,\xi)$ is the dimensionless low-frequency density fluctuation associated with the ion-acoustic field, $\tau$ is the dimensionless time, $\alpha$ is a real constant. The relation between dimensionless and dimension variables is:

$$\xi = \frac{2}{3}\sqrt{\mu}\frac{\mathbf{r}}{d_e}, \tau = \frac{2}{3}\mu\omega_{pe}t, \mu = \frac{m_e}{m_i}, \alpha = \frac{c^2}{3v_{Te}^2},$$

$$\mathbf{E}(\xi,\tau) = \frac{\sqrt{3}\mathbf{E}(\mathbf{r},t)}{4\sqrt{\pi\mu n_0 T_e}}, \mathbf{B}(\xi,\tau) = \frac{3e\mathbf{B}(\mathbf{r},t)}{4\mu m_e c\omega_{pe}}, n = \frac{3}{4\mu}\frac{\delta n_e}{n_0}, \qquad (8)$$

where $\omega_{pe}$ is the frequency of plasma, $m_e$ is the mass of electron, $m_i$ is the mass of ion, $v_{Te}$ is the thermal velocity of the electron.

In fact, the Eqs (4), (5) and (6) are a generalization of the Zakharov equations by taking into account the self-generated magnetic field. Eq. (4) is the driven ion-sound equation. Density perturbations are driven by the (divergence of the) ponderomotive force in the right-hand side of Eq. (4); the density responses in general are governed by the linear ion-acoustic operator on the left-hand side. In the static limit, we can neglect the first term on the left-hand side of Eq. (4) and we assume that the perturbed densities vanish at the "boundary" where the envelope fields go to zero, then Eq. (4) is reduced to

$$n(\tau,\boldsymbol{\xi}) \approx -|\mathbf{E}(\tau,\boldsymbol{\xi})|^2.$$

It is obvious as seen from (8) that the high-frequency plasmons are trapped inside the local zone, where the ponderomotive force – the radiation pressure force—acts on the plasmas so as to create a density depression. Eq. (6) describes the evolution of the varying magnetic field, which is set up by bits of the high-frequency fields; as was discussed, the term in (6), $\mathbf{E}(\boldsymbol{\xi},\tau) \times \mathbf{E}^*(\boldsymbol{\xi},\tau)$, is the "spin" of the fields [17], hence in other terminology, the excited magnetic field results from the intrinsic rotation of wave field with the high frequency. And the first term in the right-hand side of Eq. (5) represents the nonlinear frequency shift in the presence of particle density fluctuations; the last term describes the phase inhomogeneity via $\mathbf{B}(\boldsymbol{\xi},\tau)$.

Now to study the stability problem on the basis of the spontaneous magnetic field equations. We superimpose a small disturbance on the initial state solution of the equations and linearize the equations with reference to the small disturbance, if and only if the amplitude of the small disturbance be amplified, instability will appear; it is called unstable in the Lyapunov sense.

Supposing the following initial state:

$$\mathbf{E}_\mathrm{I} = \mathbf{E}_0 e^{i\mathbf{k}_0\cdot\boldsymbol{\xi}-i\omega_0\tau},\ n_\mathrm{I}=0,\ \mathbf{E}_0 = Const,\ \mathbf{B}_\mathrm{I} = \mathbf{B}_0 = B_0 \mathbf{b}, \tag{9}$$

and

$$\mathbf{k}_0 \cdot \mathbf{E}_0 \propto \mathbf{k}_0 \cdot \mathbf{e}_0 = 0,$$

$\mathbf{e}_0$ is a complex unit vector satisfying $\mathbf{e}_0 \cdot \mathbf{e}_0^* = 1$, then Eqs (5) and (6) become:

$$\omega \mathbf{E}_0 = \alpha k^2 \mathbf{E}_0 - i(\mathbf{e}_0 \times \mathbf{b})EB_0, \tag{10}$$

$$(-\frac{\partial^2}{\partial \tau^2} + \nabla \times \nabla \times)\mathbf{B} = i\frac{2}{3}\nabla \times \nabla \times [\frac{\partial}{\partial \tau}(\mathbf{E}_0 \times \mathbf{E}_0^*)]. \tag{11}$$

Doting Eq.(10) by $\mathbf{e}_0^*$, Eq.(10) becomes:

$$\omega_0 = \alpha k_0^2 + \beta B_0, \beta \equiv i\mathbf{e}_0^* \cdot (\mathbf{b} \times \mathbf{e}_0) = \mathbf{b}\cdot[i(\mathbf{e}_0 \times \mathbf{e}_0^*)]. \tag{12}$$

Transforming Eq.(12) into the dimensional form, that is:

$$\omega_0^2 = \omega_{pe}^2 + c^2 k_0^2 + \Omega_e \omega_{pe},\ \Omega_e \ll \omega_0,\ \omega_0 \gg ck. \tag{13}$$

The equation Eq.(13) is the dispersion relation of transverse plasmon in magnetized plasma [18]. If the magnetic field is a steady with two-dimensional, Eq.(11) becomes

$$\frac{\partial^2}{\partial \xi_x^2} B_{0\xi_x} + \frac{\partial^2}{\partial \xi_y^2} B_{0\xi_x} = 0, \quad \frac{\partial^2}{\partial \xi_x^2} B_{0\xi_y} + \frac{\partial^2}{\partial \xi_y^2} B_{0\xi_y} = 0;$$

obviously, the following circular field is its solution:

$$B_{0\xi_x} = -B_{00} \frac{\xi_x}{\ell}, \quad B_{0\xi_y} = B_{00} \frac{\xi_y}{\ell}, \quad \xi_x^2 + \xi_y^2 \leq \ell^2. \tag{14}$$

We're going to study the modulational instabilities by the hybrid transverse plasmon (12) in the circular magnetic field (14).

### 3 Modulation instability in the circular magnetic field

Inserting

$$\mathbf{E} = \mathbf{E}_I + \delta \mathbf{E}, \quad \mathbf{B} = \mathbf{B}_0 + \mathbf{B}_{II} \tag{15}$$

into Eqs (4), (5) and (6), and neglecting the small perturbations upon to the second order, we can get linearized equations:

$$(n_{II})_{\tau\tau} - \nabla^2 n_{II} = \nabla^2 [\mathbf{E}_I \cdot (\delta \mathbf{E})^* + \mathbf{E}_I^* \cdot (\delta \mathbf{E})], \tag{16}$$

$$i(\delta \mathbf{E})_\tau - \alpha \nabla \times \nabla \times \delta \mathbf{E} = n_{II} \mathbf{E}_I + i \mathbf{E}_I \times \mathbf{B}_{II} + i \delta \mathbf{E} \times \mathbf{B}_0, \tag{17}$$

$$(\frac{\partial^2}{\partial \tau^2} + \nabla^2) \mathbf{B}_{II} = -i \frac{2}{3} \nabla \times \nabla \times [\frac{\partial}{\partial \tau} (\delta \mathbf{E} \times \mathbf{E}_I^* + \mathbf{E}_I \times \delta \mathbf{E}^*)]. \tag{18}$$

To study the instability of circular self-generated magnetic field, we can superimposed small disturbances as follows on Eqs.(16), (17) and (18):

$$n_{II} = \frac{1}{2} n [e^{i\mathbf{k} \cdot \xi - i\omega\tau} + e^{-i\mathbf{k} \cdot \xi + i\omega\tau}], \tag{19}$$

$$\mathbf{B}_{II} = \frac{1}{2} \mathbf{B} [e^{i\mathbf{k} \cdot \xi - i\omega\tau} + e^{-i\mathbf{k} \cdot \xi + i\omega\tau}], \tag{20}$$

$$\delta \mathbf{E} = [\mathbf{E} e^{i\mathbf{k} \cdot \xi - i\omega\tau} + \mathbf{E}^+ e^{-i\mathbf{k} \cdot \xi + i\omega\tau}] e^{i\mathbf{k}_0 \cdot \xi - i\omega_0 \tau}, \tag{21}$$

where $\mathbf{E}$, $\mathbf{E}^+$ is transverse perturbation:

$$\mathbf{E} = E \mathbf{e}_1, \quad \mathbf{E}^+ = E^* \mathbf{e}_2^+, \quad \mathbf{e}_1 \perp (\mathbf{k} + \mathbf{k}_0), \quad \mathbf{e}_2^+ \perp (\mathbf{k} - \mathbf{k}_0), \tag{22}$$

here $\mathbf{e}_1$, $\mathbf{e}_2$ is real unit vector.

From Eq.(16) we can get:

$$-\frac{n}{2}(\omega^2 - k^2) = -k^2 [E(\mathbf{e}_2^+ \cdot \mathbf{E}_0) + E(\mathbf{e}_1 \cdot \mathbf{E}_0^*)]. \tag{23}$$

And we can get two equations from Eq.(17):

$$(\omega_+ - \alpha k_+^2) E \mathbf{e}_1 = \frac{n}{2} \mathbf{E}_0 + \frac{1}{2} i \mathbf{E}_0 \times \mathbf{B} + i E \mathbf{e}_1 \times \mathbf{B}_0, \tag{24}$$

$$(\omega_- + \alpha k_-^2) E^* \mathbf{e}_2^+ = -\frac{n}{2} \mathbf{E}_0^* + \frac{1}{2} i \mathbf{E}_0^* \times \mathbf{B} + i E^* \mathbf{e}_2^+ \times \mathbf{B}_0, \tag{25}$$

where $\omega_\pm = \omega \pm \omega_0$, $\mathbf{k}_\pm = \mathbf{k} \pm \mathbf{k}_0$.

Combining Eqs (23), (24) and (25), we obtain:

$$n(\omega^2 - k^2) = k^2 [\frac{-n\mathbf{E}_0^* + i\mathbf{E}_0^* \times \mathbf{B} + 2E^*\mathbf{e}_2 \times \mathbf{B}_0}{\omega_- + \alpha k_-^2} \cdot \mathbf{E}_0 + \frac{n\mathbf{E}_0 + i\mathbf{E}_0 \times \mathbf{B} + 2E\mathbf{e}_1 \times \mathbf{B}_0}{\omega_+ - \alpha k_+^2} \cdot \mathbf{E}_0^*]. \quad (26)$$

Doting both sides of the Eq.(25) by $\mathbf{E}_0$, we can get:

$$\frac{n}{2}|\mathbf{E}_0|^2 = -(\omega_- + \alpha k_-^2)E^*\mathbf{e}_2^+ \cdot \mathbf{E}_0 + \frac{1}{2}i\mathbf{E}_0 \cdot (\mathbf{E}_0^* \times \mathbf{B}) + iE^*\mathbf{E}_0 \cdot (\mathbf{e}_2^+ \times \mathbf{B}_0). \quad (27)$$

Then substituting Eq.(27) into Eq.(26), we can get the dispersion equation of the circular magnetic field:

$$[-2(\omega_- + \alpha k_-^2)\mathbf{e}_2^+ \cdot \mathbf{e}_0 + ik^2 \frac{4}{3}\omega|\mathbf{E}_0|^2 G(\theta) + 2iB_0\mathbf{e}_0 \cdot (\mathbf{e}_2^+ \times \mathbf{b})](\omega^2 - k^2 + k^2|\mathbf{E}_0|^2 \chi)$$

$$= ik^2 [\frac{4}{3}\chi\omega|\mathbf{E}_0|^4 G(\theta) + |\mathbf{E}_0|^2 B_0(\frac{2\mathbf{b}\cdot(\mathbf{e}_0 \times \mathbf{e}_2^+)}{\omega_- + \alpha k_-^2} + \frac{2\mathbf{b}\cdot(\mathbf{e}_0^* \times \mathbf{e}_1)}{\omega_+ - \alpha k_+^2})], \quad (28)$$

where

$$G = \frac{E_0}{E_0^*}[(\mathbf{e}_2^+ \cdot \mathbf{e}_0) - (\mathbf{e}_2^+ \cdot \mathbf{e}_0^*)(\mathbf{e}_0 \cdot \mathbf{e}_0)] - [(\mathbf{e}_1 \cdot \mathbf{e}_0)(\mathbf{e}_0^* \cdot \mathbf{e}_0^*) - (\mathbf{e}_1 \cdot \mathbf{e}_0^*)]$$

$$+ [\mathbf{e}_k \cdot (\mathbf{e}_0 \times \mathbf{e}_0^*)][\mathbf{e}_k \cdot (\mathbf{e}_1 \times \mathbf{e}_0^*)] - \frac{E_0}{E_0^*}[\mathbf{e}_k \cdot (\mathbf{e}_0 \times \mathbf{e}_0^*)][\mathbf{e}_k \cdot (\mathbf{e}_2^+ \times \mathbf{e}_0)], \quad (29)$$

$$\chi = \frac{2\beta B_0 - 2\alpha k^2}{-\beta^2 B_0^2 + 2\beta\alpha B_0 k^2 - 2\alpha^2(k^4 - 4k^2 k_0^2) - 4\alpha\omega\mathbf{k}\cdot\mathbf{k}_0 + \omega^2}, \quad (30)$$

$\theta$ is the angle between $\mathbf{k}$ and $\mathbf{k}_0$, $\mathbf{e}_0 = \frac{\mathbf{E}_0}{E_0}$, $\mathbf{e}_0^* = \frac{\mathbf{E}_0^*}{E_0^*}$, $\mathbf{e}_1 = \mathbf{e}_{1x} + \mathbf{e}_{1y} + \mathbf{e}_{1z}$,

$\mathbf{e}_2^+ = \mathbf{e}_{2x}^+ + \mathbf{e}_{2y}^+ + \mathbf{e}_{2z}^+$.

We can describe this circular magnetic field as follows:

$$\mathbf{B}_0 = B_0\mathbf{b}_0, \quad \mathbf{b}_0 = \cos\varphi\mathbf{e}_x + \sin\varphi\mathbf{e}_z,$$

where $\cos\phi = -\frac{z}{\sqrt{x^2 + z^2}}$, $\sin\phi = \frac{x}{\sqrt{x^2 + z^2}}$, $x^2 + z^2 \leq l^2$.

Numerical calculations are made for Eq.(28), supposing $\mathbf{k}_0 = k_0\mathbf{e}_z$, when $\mathbf{k}$ is on the X-Z plane and $\mathbf{k}_0 \gg \mathbf{k}$, there is approximation $\mathbf{e}_k \perp \mathbf{e}_1$ and $\mathbf{e}_k \perp \mathbf{e}_2$. So, $\mathbf{e}_k$, $\mathbf{e}_1$ and $\mathbf{e}_2$ can be written as: $\mathbf{e}_k = \sin\theta\mathbf{e}_x + \cos\theta\mathbf{e}_z$, $\mathbf{e}_1 = -\sin\theta\cos\theta_1\mathbf{e}_x + \cos\theta_1\mathbf{e}_y + \sin\theta\cos\theta_1\mathbf{e}_z$, $\mathbf{e}_2 = -\cos\theta\cos\theta_1\mathbf{e}_x - \sin\theta_1\mathbf{e}_y + \sin\theta\cos\theta_1\mathbf{e}_z$.

Making $\mathbf{e}_0 = \frac{1}{\sqrt{2}}(\mathbf{e}_x + i\mathbf{e}_y)$, $E_0$ is real. We select the largest positive virtual root $\gamma_{max} = |Im(\omega_i)|_{max}, i=1,2,\cdots,6$ among six non-conjugate roots as growth rate of modulation instability, and the dispersion curve can be obtained.

The parameters are chosen as $I \sim 5\times10^{15}(W/cm^2)$, $T_e = 10^8 K$, $n_e = 8\times10^{20} cm^{-3}$, $|E_0|^2 = 73$, $\alpha = 20$, $k_0 = 4\pi/150$, $B_0 = 25$ (corresponding to about $0.7\,MG$), and the dispersion curve for different $\theta$ is shown in Fig.1. When $\theta = \pi/2$, the dispersion curve for different $B_0$ is shown in Fig.2.

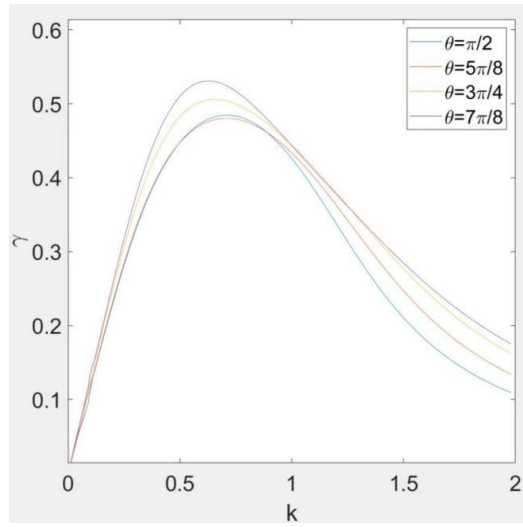

FIG.1. The curve of the dispersion relation ($|E_0|^2 = 73$, $\alpha = 20$, $k_0 = 4\pi/150$, $B_0 = 25$). The parameters of each curve are taken as follows: $\theta = \pi/2$, $\theta = 5\pi/8$, $\theta = 3\pi/4$, $\theta = 7\pi/8$.

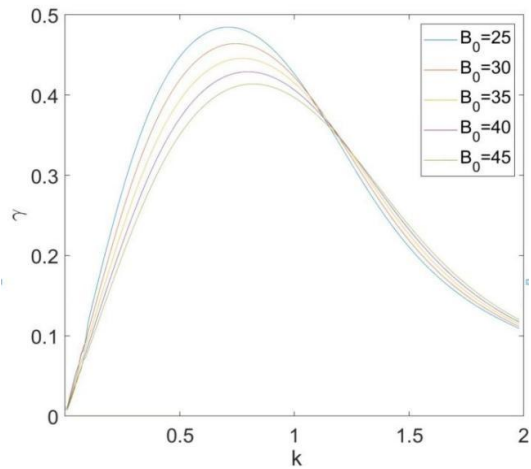

FIG.2. The curve of the dispersion relation ($|E_0|^2 = 73$, $\alpha = 20$, $k_0 = 4\pi/150$, $\theta = \pi/2$). The parameters of each curve are taken as follows: $B_0 = 25$, $B_0 = 30$, $B_0 = 35$, $B_0 = 40$, $B_0 = 45$.

## 4 Self-siminal collapse and jets for electron beam

The modulational instabilities by the magnetized plasmons, as aforesaid, make the electric and a magnetic field disturbance obtain great increment rate $\gamma_M$ (proportional to the initial limited field amplitude); it is shown that magnetic modulation is a kind of long wave instabilities with [19]

$$k_0 \ll k, \qquad (31)$$

i.e., with respect to long waves in the initial state, the modulational interactions can create an energy flow towards higher $k$; in other words, then space region of the fields ($\delta \mathbf{B}, \delta \mathbf{E}$) is compressed (localized), resulting in the enhancement of the fields and rarefaction of the density field $\delta n(t, \mathbf{r})$. Hence the nonlinear development of the modulational instability will lead to collapse.

Self-generated magnetic field equations (4)-(6) are very complicated nonlinear vector field equations, so far we have not found analytically the collapse solutions. Now let us find a self-similar solution [20]. Assuming that there is a perturbation $\delta \mathbf{E}$ at an initial time, then from the condition of transverse waves, $\mathbf{E}(\tau, \boldsymbol{\xi}) = 0$, we have $\mathbf{k}_0 \cdot \delta \mathbf{E} = 0$. That is, the collapse in longitudinal perturbation (parallel to the $\mathbf{k}_0$) is not permitted. Hence the collapse is anisotropic. Now let us assume that the nonlinear entity has a form with a scale $\xi$ and a longitudinal scale $L$, $L \gg \xi$; and $\mathbf{E}$ has characteristic time-space scales ($\tau, \xi$). For sub-sonic motion ($\xi/\tau < 1$), it is possible to neglect the term ($\partial^2/\partial \tau^2$) in Eqs (4) and (6). Then one has $n \sim - |\mathbf{E}|^2$ and $\mathbf{B} = i \frac{2}{3} \frac{\partial}{\partial \tau} (\mathbf{E} \times \mathbf{E}^*)$. Then we write the order of magnitudes of terms in Eq. (5):

$$\frac{1}{\tau} \approx c_1 \frac{\alpha}{\xi^2} + c_2 |\mathbf{E}|^2 + c_3 \frac{|\mathbf{E}|^2}{\tau}, \qquad (32)$$

where we have used $\mathbf{E} \quad \mathbf{B} \sim \frac{1}{\tau} \mathbf{E}$ $(\mathbf{E} \quad \mathbf{E}^*) \sim \frac{1}{\tau} \mathbf{E} |\mathbf{E}|^2$; $|c_1|$, $|c_2|$ and $|c_3|$ are constants of order unity; and $\tau = \tilde{t}_0 - \tilde{t}$, $\tilde{t}_0$ corresponds to $\bar{W}^p \geq 1$, at which the perturbed fields become too strong for Eqs (4), (5) and (6) to be valid. Thus we obtain from Eq. (32) for the sub-sonic collapse

$$|\mathbf{E}|^2 \sim \frac{1}{\tau} \qquad (33)$$

provided that the first and the last terms on the right side of Eq. (32) are balanced, this implies

$$\alpha \frac{\tau}{\xi^2} \sim |\mathbf{E}|^2,$$

or in view of Eq. (33),

$$\xi \sim \tau.$$

In addition, multiplying (5) by $\mathbf{E}^*$ and subtracting from it the complex conjugate equation multiplied by $\mathbf{E}$, we get

$$i\frac{\partial}{\partial \tau}|\mathbf{E}|^2 + [\alpha(\nabla \mathbf{E}^*)\nabla\mathbf{E} - \alpha(\nabla \mathbf{E})\nabla\mathbf{E}^*] = 0 ;$$

then integrating over all space yields the conservation of the plasmon number:

$$\int |\mathbf{E}|^2 \, d\xi \, dL = \text{const.},$$

this gives $|\mathbf{E}|^2 \pi \xi^2 = \text{const.}$, i.e.,

$$\frac{\xi^2}{\tau} = Const..$$

Therefore, a self-similar collapse solution can be asymptotically shown as:

$$|\mathbf{E}| \simeq \frac{1}{\sqrt{\tau}} F(\frac{\xi^2}{\tau}), \tag{34}$$

where F is a function determined by the initial conditions. This solution is referred to as "similar" because the initial shape in space $F(\frac{\xi^2}{\tau})$ remains unchanged in evolution with time, although it is compressed and narrowed due to replacing $\xi$ with $\xi^2$.

But it is impossible to stop collapse in the stage of sub-sonic motion, because nonlinear entity with three dimensional cannot establish a balance between disturbance hot pressure (it makes plasma go into entity) and wave pressure (it makes plasma drive out of entity), local field strengths increase with the collapse, eventually led to the supersonic movement. For supersonic motion ($\xi/\tau = \text{const.} > 1$), it is possible to neglect the term ($\nabla^2$) in Eqs (4) and (6); and by similar considerations, it is proved that the solution (34) still is valid. Then the maximum growth rate $\gamma_{coll,\max}$ by collapse motion could be obtained from the self-similar solution (33) as:

$$\frac{\xi_M^2}{\tau_M} = \frac{\xi_{coll}^2}{\tau_{coll}}, \quad (\xi_M = k_M^{-1}) \quad . \tag{35}$$

We may identify the $\xi_{crit}$ as the scale to stop collapse, which corresponds to $\bar{W}^p \sim 1$.

On the other hand, it should be pointed out that at this time we treat the modulation instability for monochromatic wave, which is a kind of zero-threshold instability. However, in fact, envelope fields, the turbulent fields, shown as (3c), have a broad wavepacket with a spread $\Delta k$ of wavenumber and frequency width $\Delta \omega$. Obviously, the conditions for the instability not to be affected are [17]

$$\Delta k \ll k_{\max}, \quad \Delta \omega \ll \gamma_M ; \tag{36}$$

as above, the fields are compressed and narrowed, that is, $\Delta k$ ($\sim 1/\xi$) is increased; and as $\Delta k \sim k_{\max}$, the conditions for the threshold (36) will not be fulfilled. Then the suppression of the instability occurs [19]. Hence in the case one has $1/\xi_{crit} \sim (\Delta k)_{\max} \sim k_{\max}$. Therefore, we obtain eventually

$$\tau_{coll} \approx \tau_M, \quad \xi_{crit} \approx 1/k_{\max}. \tag{37}$$

The zeroth-order moment of Vlasov equation (1) is the continuity equation for charged particles, such as electron beam, and it can be found by integration of the Vlasov equation over the entire velocity space,

$$\frac{\partial n_e(t,\mathbf{r})}{\partial t} + \nabla \cdot n_e(t,\mathbf{r})u_e(t,\mathbf{r}) = 0 .\tag{38}$$

According to the in front of the set,

$$n_e = n_0 + \delta n, \quad \mathbf{u}_e = 0 + \mathbf{u}_e ,$$

and omitting small disturbances with the second-order one gets from Eq. (31)

$$\frac{\partial}{\partial t}\frac{\delta n}{n_o} = -\nabla \cdot \mathbf{u}_e(t,\mathbf{r}) \sim -\frac{1}{L'}u_e .$$

On the other hand, due to the large electrical conductivity in plasma, freezing effect makes charged particles can't cross the compression ring area, so that, $\frac{1}{L'}u_e \approx k'_L \mathbf{u}_{z,e}$; here $L'$ is a longitudinal characteristic scale over which amplitudes of the fields, involving fluid field $u_{z,e}$, change obviously. So we may be able to identify $L'$ as $\lambda_0 = 2\pi / k'_0$ with $k'_0 = \frac{2}{3}\sqrt{\mu}\frac{1}{d_e}k_0$. Then by use of Eq. (4)

$$\frac{\delta n}{n_0} \sim \frac{4\mu}{3}\frac{\tau_{coll}^2}{\xi_{crit}^2}\left|\mathbf{E}(\tau,\boldsymbol{\xi})\right|^2_{max} ,$$

one has

$$(u_{z,e})_{coll} \sim \frac{2\pi}{k'_0}\gamma_{t,max}\frac{4\mu}{3}\frac{k_{max}^2}{\gamma_{max}^2}\left|\mathbf{E}(\tau,\boldsymbol{\xi})\right|^2_{max} ;$$

and the collapse induced by the modulational instability must be terminated in the condition (2), i.e.,

$$\frac{8}{3}\mu\left|\mathbf{E}(\tau,\boldsymbol{\xi})\right|^2_{max} = \frac{\left|\mathbf{E}^t(t,\mathbf{r})\right|^2}{8\pi n_0 k_B T_0} < 1 ,$$

or,

$$\left|\mathbf{E}(\boldsymbol{\xi},\tau)\right|^2_{max} \sim \frac{3}{8\mu}\chi_N ,\tag{39}$$

where, $\chi_N < 1$.

Hance,

$$(u_{z,e})_{coll} \sim 2\pi\frac{k_{max}}{k_0}\frac{1}{2}\frac{\frac{2}{3}\mu\omega_{pe}\gamma_{max}}{\frac{2}{3}\sqrt{\mu}k_d k_{max}}\chi_N\frac{k_{max}^2}{\gamma_{max}^2} \sim \pi\frac{k_{max}}{k_0}\sqrt{\mu}\frac{k_{max}}{\gamma_{max}}\chi_N \mathrm{v}_{Te} ,\tag{40}$$

where $k_{max}$ is the wave-number at which the maximum growth rate of the perturbation fields occurs.

If $\alpha = 20$, $\left|\mathbf{E}_0\right|^2 = 73$, $T_e = 10^8 (K)$, $B_0 = 25$, $\theta = \frac{\pi}{2}$, $\frac{k_{max}}{k_0} \equiv \eta \sim 40$, $\chi_N \sim 0.37$,

$\frac{k_{max}}{\gamma_{max}} \sim 1.42$ (see Fig. 1), then the velocity of electron jet is $(u_{z,e})_{coll} \sim 5.5 \times 10^9 \text{cm/s}$ (corresponding to about $20 \text{keV}$), which is in the range of the experimental results in Y. T. Li *et al.* [4].

## 5 Conclusions

This paper is concerned with the circular self-generated magnetic field in laser plasma and the electron jet generated by the magnetic instability. From the above study, we arrive at the following conclusions:

(1) From Fig.1, one can find that when $\theta = \pi/2$, it is a vertical disturbance ($\mathbf{k} // \mathbf{k}_0$), the maximum growth rate is minimal. And the maximum growth rate of modulation instability becomes larger when $\theta$ increases. It was found by dispersion equation that magnetic field responds more to the axial perturbations. The circular self-generated magnetic field has little effect on radial electric field.

(2) From Fig.2, the growth rate of modulation instability becomes smaller as $B_0$ increases. This indicates that stronger circular self-generated magnetic fields interact with more particles in plasma, and inhibit the nonlinear interaction of wave and wave in plasma.

(3) Higher the magnetic field is, bigger the energy of electron jet. Our calculations show that the electron jet generated by the magnetic instability is strong enough. This is expected as that a stronger self-generated magnetic field increase the speed of the electron jet.

## Acknowledgments

This work is supported by the National Natural Science Foundation of China (12065018) and the Fund from the Jiangxi Provincial Key Laboratory of Fusion and Information Control (No. 20171BCD40005).

## Author contributions

All the authors were involved in the preparation of the manuscript. All the authors have read and approved the final manuscript.

## Data Availability Statement

This manuscript has no associated data or the data will not be deposited. [Authors' comment: All data is included in the manuscript].